\documentclass[useAMS,usenatbib]{mn2e}  

\def\lsim{~\rlap{$<$}{\lower 1.0ex\hbox{$\sim$}}}
\def\gsim{~\rlap{$>$}{\lower 1.0ex\hbox{$\sim$}}}

\def\e{$\pm$} 
\newcommand{\ltsima} {$\; \buildrel < \over \sim \;$} 
\newcommand{\simlt} {\lower.5ex\hbox{\ltsima}} 
\newcommand{\gtsima} {$\; \buildrel > \over \sim \;$} 
\newcommand{\simgt} {\lower.5ex\hbox{\gtsima}} 
\def\kms{km$\,$s$^{-\!1}$} 
\def\vsi{$v\: \sin i$}

\title[Rotational velocities in symbiotic stars - III]
{Rotational velocities of the giants in symbiotic stars: \\ 
III. Evidence of fast rotation in S-type symbiotics
\thanks{based on: observations obtained in ESO
 programs 073.D-0724A  and 074.D-0114,
data from the UVES Paranal Observatory Project (ESO DDT Program ID 266.D-5655),
spectral data retrieved from the ELODIE archive at Observatoire de Haute-Provence (OHP)}
\thanks{e-mail: rkz@astro.bas.bg  mfb@astro.livjm.ac.uk  andreja.gomboc@fmf.uni-lj.si
}} 
\author[Zamanov, Bode, Melo, et al. ] {
R. K. Zamanov$^{1}$,  
M. F. Bode$^{2}$,
C. H. F. Melo$^{3}$, I. K. Stateva$^{1}$, R. Bachev$^{1}$,\\
\\ 
\LARGE 
{\rm  A. Gomboc$^{4}$, R.~Konstantinova-Antova$^{1}$,
K. A. Stoyanov$^{1}$} \\
\\
$^1$ Institute of Astronomy, Bulgarian Academy of Sciences, 
       72 Tsarighradsko Shousse Blvd., 1784 Sofia, Bulgaria \\ 
$^2$ Astrophysics Research Institute, Liverpool John Moores University,  Twelve Quays House, 
     Birkenhead, CH41 1LD, UK\\ 
$^3$ European Southern Observatory, Casilla 19001, Santiago 19, Chile \\
$^4$ Department of Physics, University of Ljubljana, 
       Jadranska 19, 6100 Ljubljana, Slovenia
}  
\begin{document} 
\date{Accepted . Received 2008 May 26; in original form 2007 December 29} 
 \pagerange{\pageref{firstpage}--\pageref{lastpage}} \pubyear{2008} 
 
\maketitle 
 
\label{firstpage} 
 
\begin{abstract}  
We have measured  the  projected rotational 
velocities (\vsi) in a number of symbiotic stars and  M giants 
using high resolution spectroscopic observations. 
On the basis  of our measurements and data from the literature, 
we  compare the rotation of mass-donors in symbiotics  with \vsi\ of field giants
and  find  that: 
(1) the K giants in S-type symbiotics rotate at \vsi$\: >\!4.5$~\kms, which is
 2-4 times faster than the field K giants;
(2) the M giants in S-type symbiotics rotate on average 1.5 times faster than 
the field M giants. Statistical tests show that these differences are highly 
significant -- p-value $< 10^{-3}$  in the spectral type bins K2III-K5III, M0III-M6III, 
and M2III-M5III;
(3) our new observations of  D'-type symbiotics also confirm that they are fast rotators.

As a result of the rapid rotation, the cool giants in symbiotics should have 
3-30 times larger mass loss rates. Our results suggest also that bipolar ejections
in symbiotics seem to happen in objects where the mass donors 
rotate faster than the orbital period.  

All spectra used in our series of  papers can be obtained upon request 
from the authors. 
\end{abstract} 
 
 
\section{Introduction} 
The Symbiotic stars (SSs)  are interacting binaries
consisting of a cool giant and a hot companion. In most cases 
the hot component is a white dwarf (WD), but it can also be a 
main sequence star, hot subdwarf, or neutron star 
(see Kenyon 1986, Corradi et al. 2003).
The symbiotic phase represents a late stage in stellar evolution 
and about 200 such objects are known (Belczy{\' n}ski et al. 2000).  
On the basis of their IR properties, SSs have been classified 
into stellar continuum  (S) and dusty (D or D') types (Allen 1982). 
The D--type systems contain Mira variables as mass donors.  
D'-type symbiotics are characterized by an earlier spectral type giant 
(F-K) and lower dust temperatures. 
Soker (2002) has predicted theoretically that the cool companions in symbiotic systems  
are likely to rotate much faster than isolated cool giants or those in wide binary 
systems.

This is the third in a series of papers exploring the rotational velocities  
of the mass donating (cool) components of SSs.                           
In this paper we present \vsi\ measurements and expand our sample of SSs. We then 
perform comparative analyses and  explore theoretical predictions 
that the mass donors in S-type symbiotics  
are faster rotators compared with  field giants.

\section{Projected rotational velocity}

\subsection{Our observations}

\begin{table*}
\caption{Journal of observations and projected rotational velocities. 
 In the table are given as follows: 
 the name of the object, the IR type, date of observation (YYYY-MM-DD), 
 the modified Julian Date (JD - 2400000.5) of the start  of the observation,
the spectral type.
\vsi\  is the projected rotational 
velocity of the cool giant as measured with FWHM and CCF  methods.
If other measurements of \vsi\ exist, they are also given
in the last column.}  
\begin{tabular}{lllrrrrrrrr}
\hline
 object        &IR & Date-obs    &   MJD-obs   & Sp      &    \vsi  & \vsi   & other  &  \\
               &   &             &             &         &   FWHM   &  CCF   &        &  \\
               &   &             &             &         &  [\kms]  & [\kms] & [\kms] &  \\
\hline
               &   &             &             &         &                               \\           
   AG Peg      & S & 2004-10-02  & 53280.0676  & M4III   &  8.5\e1.5    &  7.5\e1.5    &  4.5\e1$^a$ \\
   BX Mon      & S & 2004-11-05  & 53314.2610  & M5III   & 11.0\e1.5    &  9.4\e1.5    &  6.8\e1$^a$ \\
   Hen 3-905   & S & 2005-02-02  & 53403.1847  & M3III   &  9.2\e1.5    &  7.3\e1.5    &  \\
   LIN 358     & S & 2004-10-13  & 53291.0994  & K5III   &  4.8\e1.5    &  5.9\e1.5    &  \\
   PN Sa 3-22  & S & 2005-02-01  & 53402.2405  & M4.5III &  9.7\e1.5    &  9.8\e1.5    &  \\
   V694 Mon    & S & 2004-11-05  & 53314.3158  & M6III   & 10.3\e1.5    &  -- -- --    &  \\
   V840 Cen    &-- & 2005-02-01  & 53402.2858  & K5III   & 31.5\e3.0    &  24.4\e3.0   &  \\
    &  &  \\ 
   WRAY 15-157 & D'& 2004-11-06  & 53315.3276  & G5III   & 44\e5        &  37\e5               & \\
   AS201       & D'& 2004-11-14  & 53323.3309  & F9III   &  27\e3       &  29.3\e3 & 25\e5$^b$ & \\  
  \hline
  \label{tab-log}                                                      
  \end{tabular}        
\\                                              
$^a$Fekel et al. (2003),
$^b$Pereira et al. (2005).
 \end{table*}	  
                
The new data have been obtained  from  2004 October to  2005 February
with the same set-up -- the FEROS spectrograph at the 2.2m telescope 
of ESO (La Silla). 
These are S- and D'-type SSs with $0^h<RA<12^h$, 
and catalogue magnitude brighter than $V<12.5$.

The details of the observational set-up, data reduction and the techniques of \vsi\
measurements are given in Paper~I and II (Zamanov et al. 2005, 2006). 
Here we also apply two techniques of \vsi\  measurements: 
FWHM  and Cross-Correlation Function (CCF). The description of the methods 
is given in Sect.~3 of Paper II.

The new data are presented in Table~\ref{tab-log}. Here the IR type of the objects and 
the spectral types of the mass donors  are taken from the catalogue  of
Belczy{\' n}ski et al. (2000). This catalogue lists the spectral type 
of LIN~358 as a mid-K giant; we assume that it is K5III.
For V694~Mon the CCF method does not give a meaningful result, due to 
the highly complicated emission/absorption lines  in the spectrum. 
For this object we will use the result of the FWHM method,
for all other objects the CCF measurements are more reliable.

Combining  the present set  with data from Papers~I and II,
and from the literature, we have \vsi\ measured
for all southern  S- and D'-type SSs from the Belczy{\' n}ski et al. (2000) 
catalogue with $V<12.5$.
This sample should have no biases in 
the rotational speed of the cool giant, even though the sample is flux limited. 

\subsection{Catalog data}

The  giants of S-type SSs with measured \vsi\ are of spectral 
types from K2III to K5III and from M0III to M6III. 

We searched for published data of \vsi\ of  giants of similar spectral type
and found totally 362 giants in the interval 
K2-K5III  in de~Medeiros and Mayor (1999) and 
Massarotti et al. (2008).
In the interval M0III-M6III we found 
10 objects in  Glebocki et al. (2001), 4 in H{\"u}nsch et al.(2004)
and 15  in  Massarotti et al. (2008).

\subsection{Archive spectra}

\begin{table}
\caption{Projected rotational velocities of M giants, based on archive spectra 
from UVES and ELODIE archives. 
 In the table are given as follows: 
 the name of the object, the spectral type,  the projected rotational 
 velocity  (\vsi) as measured with FWHM and CCF methods with the corresponding errors.}  
\begin{tabular}{llrrrrr}
\hline
 object         & Sp     &    \vsi   &   \vsi   & \\
                &        &    FWHM   &   CCF    & \\
		&	 &  [\kms]   &  [\kms]  & \\
\hline
UVES & \\
  Menkar       &  M1.5IIIa  & 2.5\e1.5 &  2.8\e1.5  & \\
  HD 102212    &  M1III	    & 1.0\e1.0 &  1.1\e1.4  & \\
  HD 119149    &  M1.5III   & 2.3\e1.0 &  1.9\e1.5  & \\
  HD 123934    &  M1III	    & 1.0\e1.0 &  2.3\e0.9  & \\
  HD 210066    &  M1III	    & 1.0\e1.0 &  0.6\e2.8  & \\
  HD 92305     &  M0III	    & 1.7\e1.5 &  1.2\e0.7  & \\
  HD 219215    &  M2III	    & 1.0\e1.0 &  0.0\e1.0  & \\
  HD 120052    &  M2III	    & 1.5\e1.5 &  1.0\e1.5  & \\
  HD 11695     &  M4III	    & 2.5\e1.0 &  3.9\e0.9  & \\
  HD 118767    &  M5III	    & 7.5\e1.5 &  7.7\e1.0  & \\  
  HD 123214    &  M4III	    & 3.0\e1.0 &  4.4\e0.7  & \\
  HD 130328    &  M3III	    & 5.5\e2.0 &  8.2\e1.7  & \\
  HD 189124    &  M6III	    & 5.3\e1.0 &  9.1\e0.7  & \\ 
  HD 189763    &  M4III	    & 1.5\e1.5 &  4.5\e1.9  & \\
  HD 213080    &  M4.5III   & 1.5\e1.5 &  4.5\e0.8  & \\
  HD 214952    &  M5III	    & 2.5\e1.5 &  6.3\e2.8  & \\
  HD 224935    &  M3III	    & 2.5\e1.5 &  3.9\e0.6  & \\
\\
 ELODIE  & \\
  HD 006860     &  M0III    &  1.5 \e 1.5 &  5.6\e2.0   & \\   
  HD 018191     &  M6III    & 12.0 \e 2.0 &  9.6\e2.0   & \\   
  HD 042787     &  M2III    &  2.0 \e 1.5 &  5.2\e2.0   & \\   
  HD 046784     &  M0III    &  2.8 \e 1.5 &  5.4\e2.0   & \\   
  HD 097778     &  M3III    &  5.5 \e 1.5 &  7.6\e2.0   & \\   
  HD 167006     &  M3III    &  5.5 \e 1.5 &  5.2\e2.0   & \\   
  HD 168720     &  M1III    &  5.0 \e 1.0 &  5.2\e2.0   & \\   
  HD 184786     &  M4.5III  &  6.5 \e 2.0 &  7.8\e2.0   & \\   
  HD 216131     &  M2III    &  1.0 \e 1.0 &  2.5\e2.0   & \\   
  HD 218329     &  M2III    &  3.8 \e 1.5 &  5.0\e2.0   & \\   
  HD 219734     &  M2III    &  5.0 \e 1.0 &  4.9\e2.0   & \\   
 \\
 \hline
  \label{t-UVES}                                                      
  \end{tabular}        
\\                                              
 \end{table}	  

Because we have found only 29 field M giants with measured \vsi, 
we searched in the archives for spectra of red giants in the interval 
M0III-M6III. 

From  the UVES Paranal Observatory Project (Bagnulo et al., 2003),
we downloaded spectra of 17  M giants and from the ELODIE archive at Observatoire 
de Haute-Provence 
(Moultaka et al. 2004) we took  spectra of 11 more objects. 
We applied the FWHM and CCF methods to measure the \vsi\ parameter in each case. 
The results are summarized in Table~\ref{t-UVES}.

For the CCF measurements, 
the calibration of UVES values comes from an unpublished paper 
(Melo et al. in preparation). 
For ELODIE, we used the same technique as usual but using a 
$\sigma_0=4.9$~\kms, taken as the average value of Delfosse et al. (1998). 
This gives an error of $\pm2.0$~\kms\ in the \vsi.

\section{Results}

 \begin{figure}
 \mbox{} 
 \vspace{7.0cm}  
 \includegraphics{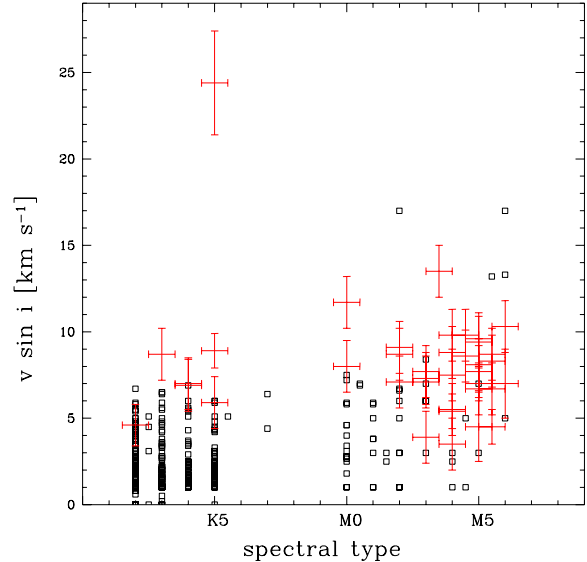}   
 \caption[]{
 Projected rotational velocity (\vsi)  versus the spectral 
 type of the giant in the spectral interval K2III-M6III.  
 S-type SSs  are indicated with error bars.
 The field giants are plotted with squares.
 }		    
\label{fig-all}     
\end{figure}	      

In Fig.\ref{fig-all}, the projected rotational velocity (\vsi)
is plotted  versus the spectral type of the giant.
The error in the spectral type of SSs is adopted as $\pm0.5$.
In Fig.\ref{fig-all}, it is seen that in the most cases the mass donors in SSs rotate 
faster than the field  giants. To test
the visual impression, as a mathematical
(statistical) approach we use  the 
Kolmogorov-Smirnov and  Mann-Whitney U-tests.
These statistical tests are performed in the
spectral type bins K2III-K5III, M0III-M6III, M2III-M5III. 
These bins are selected following the distribution of symbiotics with known \vsi, 
and avoiding the interval K6III-K9III, where we do not find symbiotics.
It deserves noting that Keenan \& McNeil (1989) do not consider K6III-K9III
as full spectral subclasses, therefore the lack of objects.

\subsection{ Bin K2-K5 giants}

 \begin{figure}
 \mbox{} 
 \vspace{8.5cm}  
 \includegraphics{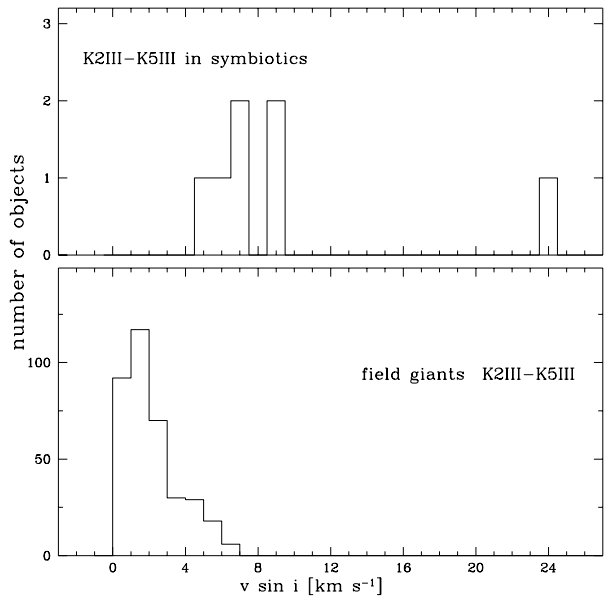}   
 \caption[]{The histogram of \vsi\  for K2-K5 giants in 
symbiotics (upper panel) compared to 
the relevant distribution for field giants of the same spectral 
classes (bottom panel). 
}		    
\label{figK}     
\end{figure}	      

The histograms for K giants are presented in Fig.~\ref{figK}. 
From  Fig.\ref{fig-all} and Fig.\ref{figK}, it is seen
that the field  giants and the giants in S--type 
symbiotic systems occupy different areas of 
\vsi\  values and that the K-giants in S--type SSs rotate 
faster than field giants. 

For the field  K giants typically \vsi$<3$~\kms. For K giants in symbiotics
$4.5\le$\vsi$\le25$~\kms. 

For 363 field  K2III-K5III giants we calculate a
mean \vsi$=$2.2~\kms,  median \vsi$=$1.8~\kms, and
standard deviation of the mean $\sigma=1.45$~\kms.

For 7  K2III-K5III giants in symbiotics,
we get a mean \vsi$=9.5$~\kms,    median \vsi$=$7.0~\kms,
and standard deviation of the mean $\sigma=$6.7~\kms.
If we exclude V840~Cen  from the sample 
we calculate a mean \vsi$=7.0$~\kms,    median \vsi$=$7.0~\kms,
and standard deviation of the mean $\sigma=$1.6~\kms.
On average SSs rotate $2-4$ times
faster than the field K giants.

The Kolmogorov-Smirnov test gives a probability of only $7.10^{-6}$
(KS statistics =0.90) that both distributions arise from the same parent population.

The comparison of the medians (Mann-Whitney U-test) 
for the symbiotic and field giants 
gives a  probability  of the  median \vsi\ of the symbiotic giants being higher 
than that of the field giants  $>0.99999$ (U~statistics~$=1659$). 

It is therefore statistically  certain that 
the K giants in symbiotics rotate faster than the field  K giants.
It has to be noted that the symbiotic V840~Cen rotates unusually fast (see Table~1)
and it is the most rapid rotator among  giants in the bin K2III-K5III
known until now.

\subsection{Bin M0-M6 giants}
 
For 53  field  M0III-M6III giants we calculate a
mean \vsi$=$4.8~\kms,  median \vsi$=$3.8~\kms, and
standard deviation of the mean $\sigma=3.6$~\kms.

For 32  M0III-M6III giants in symbiotics,
we get a mean \vsi$=9.2$~\kms,   median \vsi$=$8.0~\kms,
and standard deviation of the mean $\sigma=$8.1~\kms.

The fastest rotator among M giants is the symbiotic star
Hen~3-1674, which rotates almost at break-up velocity (see Paper II).
This object in not plotted on the figures because it is too far from the other values.

\begin{figure}
 \mbox{}   
  \vspace{8.5cm}  
 \includegraphics{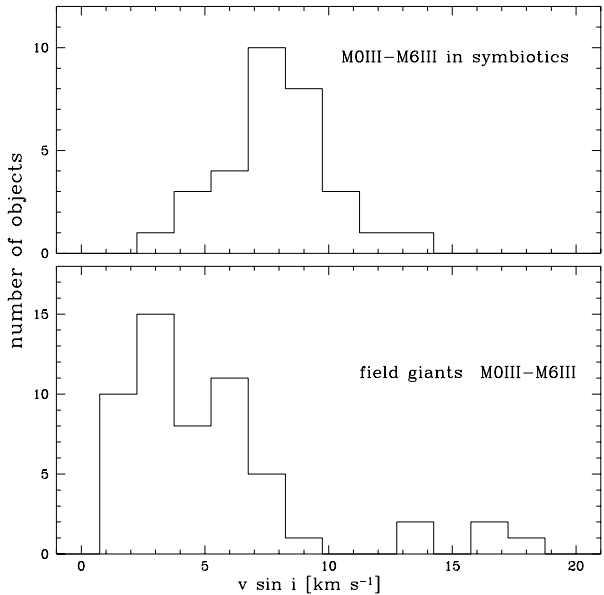}   
  \caption[]{The distribution of \vsi\ for  M0-M6  giants in 
 symbiotic binaries (upper panel) compared to the 
 distribution for the  field  M giants (bottom panel). 
  }		    
  \label{figM06}     
  \end{figure}

The distributions of \vsi\ in this bin are plotted in
Fig.\ref{figM06}.
The Kolmogorov-Smirnov test gives a probability of only $3.5\times10^{-7}$
(KS statistics =0.60) that both distributions arise from the same parent population.

The comparison of the medians (Mann-Whitney U-test) 
for the symbiotic and single giants 
gives a  probability of the  median \vsi\ of the symbiotic giants being higher 
than that of the field giants  $0.999994$ (U~statistics~$=273$). 

In the above statistical test we used the FWHM measutements of the UVES 
and ELODIE spectra. If in the statistical test we use the CCF measurements,
we obtain very similar results. The KS-test gives a probability $5\times10^{-6}$ 
(KS statistics =0.57) and  the U-test gives  0.9999999 (U~statistics~$=353.5$).

\subsection{Bin M2-M5 giants}

 \begin{figure}
 \mbox{}   
  \vspace{8.5cm}  
 \includegraphics{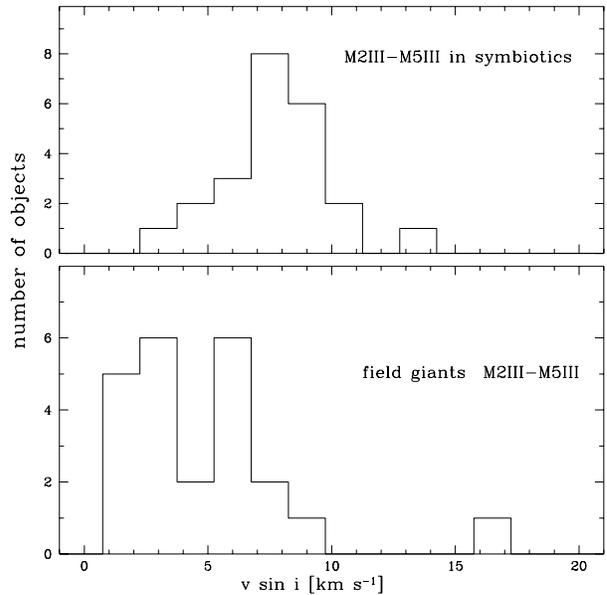}   
  \caption[]{The distribution of \vsi\ for  the bin M2III-M5III
 in  symbiotic binaries (upper panel) and for the 
 field  giants (bottom panel). 
  }		    
  \label{figM25}     
  \end{figure}	      

This bin is selected because (1) it is in the mid of symbiotic M giants, and
(2) we have  23 symbiotics and 23 field giants inside. 
The histograms for M2III-M5III giants are presented in Fig.~\ref{figM25}.
For the 23 field giants in this bin with known \vsi, we calculate 
mean \vsi$=4.8$~\kms, median \vsi$=5.0$~\kms,  
and standard deviation  of the mean $\sigma=3.55$~\kms.

For  23  M giants in symbiotics we get a
mean \vsi$=7.7$~\kms, median \vsi$=7.7$~\kms,
and standard deviation of  the mean $\sigma=2.2$ \kms.

The Kolmogorov-Smirnov test gives a probability of $1.1\times10^{-4}$ 
(K-S statistics = 0.65) that both 
distributions are extracted  from the same parent population. 
The U-test gives a probability of 0.99995 (U~statistics$=442.5$),  
for the hypothesis that the median \vsi\  of the symbiotic M-giants 
is higher than the median of the single M-giants. 

If in the statistical tests we use CCF measurements for UVES and ELODIE spectra,
we obtain KS-statistics=0.52, probability $10^{-9}$, and 
U-statistics$=134.5$, probaility 0.999999. 

\subsection{Additional checks}
 
 \begin{figure}
 \mbox{}   
  \vspace{8.5cm}  
 \includegraphics{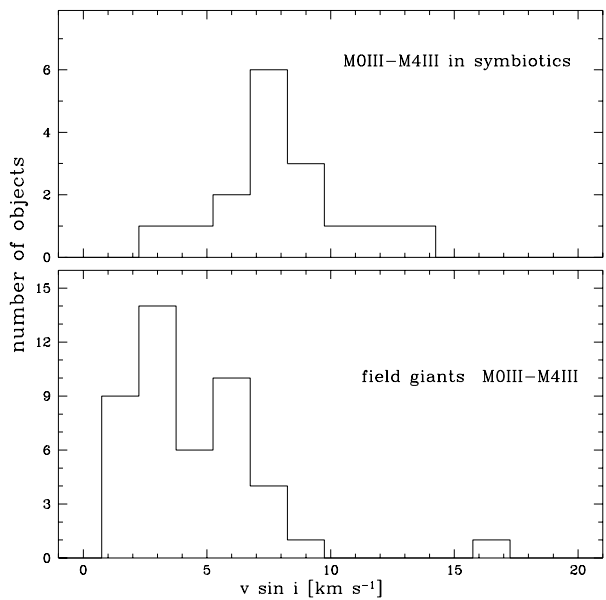}   
  \caption[]{The distribution of \vsi\ for  the bin M0III-M4III
 in  symbiotic binaries (upper panel) and for the 
 field  giants (bottom panel). 
  }		    
  \label{figM04}     
  \end{figure}	      
From Fig.\ref{fig-all}, it is seen that 
in the M0III spectral class the two fastest rotators are SSs.

As an additional check we have performed statistical test 
in bin M0III-M4III, where we have 16 symbiotics and 45 field giants inside. 
The histograms for M0III-M4III giants are presented in Fig.~\ref{figM04}.
The Kolmogorov-Smirnov test gives a probability of $1\times10^{-9}$ 
(K-S statistics = 0.64) that both 
distributions are extracted  from the same parent population. 
The U-test gives a probability of  0.999999 (U~statistics$=117.5$),  
for the hypothesis that the median \vsi\  of the symbiotic M-giants 
is higher than the median of the single M-giants. 
The statistical tests in M0III-M3III (10 SSs and 45 field giants) give
$2\times10^{-5}$ (K-S statistics = 0.83) and 0.999977 (U=38.0).
In bin M4III-M6III (21 SSs and 12 field giants) the tests 
do not give statistically significant difference: 0.1 (KS-statistics=0.44)
and 0.85 (U~statistics$=98$).

\begin{table}
\caption{Projected rotational velocities of M giants.
 In the table are given as follows: the spectral type,  
 the  mean projected rotational  velocity (\vsi, in \kms), 
 standart deviation of the mean ($\sigma$, in \kms), the number of objects.
 In the second column are given the values for the field M giants, 
 in the third - for the symbiotic stars. }
\begin{tabular}{rrrrrr}
\hline
Sp           & {\bf field giants}   & {\bf symbiotics}   & \\
             & mean\e $\sigma$ (N) & mean\e $\sigma$ (N) & \\ 

M0-M1   III  &	3.7\e1.9(23) &	9.9\e2.6(2)  & \\
M1.5-M2 III  &	4.8\e4.1(14) &	8.3\e1.1(3)  & \\
M2.5-M3 III  &	5.5\e2.0(8)  &	6.5\e1.8(4)  & \\
M3.5-M4 III  &	2.2\e1.0(3)  &	7.7\e3.3(7)  & \\
M4.5-M5 III  &	5.5\e4.0(5)  &	7.9\e1.7(9)  & \\
M5.5-M6 III  & 12.1\e5.1(4) &	7.6\e2.0(6)  & \\
 \\
 \hline
 \label{t-mean-sig}                                                      
 \end{tabular}        
\\                                              
 \end{table}	  

The mean values of \vsi\ for the M giants are presented in 
Table~\ref{t-mean-sig}. All but one of the mean \vsi\ values of SSs 
are higher than those of the field giants. 
The majority of the field M giants rotate at about \vsi\ $\sim1-6$ \kms,
while the symbiotic M giants rotate at  \vsi\ $\sim 4 - 14$ \kms

Our calculations presented above show that 
the M giants in symbiotics rotate faster
(on average 1.5 times faster) than the field  M giants, 
at least for the spectral interval from M0III to M5III. 

\subsection{D'-type symbiotics}

There are 2 D'-type SSs among our new observations.
For AS~201 our value of \vsi\ confirms the measurement 
of Pereira et al. (2005). The object is discussed in more detail in Paper~I.

WRAY~15-157 is of spectral class G5III.
The  catalogue of rotational  velocities for evolved stars (de Medeiros \& Mayor  1999)  
lists 18  objects from  spectral type G5III.  
They all rotate with \vsi$<15$ \kms.
WRAY~15-157 with \vsi$=37\pm5$ \kms, is an extremely  fast rotator for this spectral class.

There are  7 D' SSs listed in the catalogue of Belczy{\' n}ski et al. (2000).
Rotational velocities are  measured for all six southern objects.
In  Paper~I, we have shown that 4 out of 5 D'--type symbiotics
rotate rapidly. Here  we strengthen
our result: 5 out of the six southern D'--type SSs are very fast rotators. 
Furthermore, four  of them are the fastest rotators in their spectral class.

\section{Discussion}

\subsection{Reasons for the fast rotation}
Soker (2002) discussed in detail three possible mechanisms for 
speeding-up the rotation of the  mass donors in SSs:
{\bf (i)}  synchronization;
{\bf (ii)}  accretion during the main sequence phase of what is now the red giant;
{\bf (iii)} backflowing material. 

In the light of our results (see Paper II), 
synchronization plays an important role. Other mechanisms 
that can also operate in single giants as well as in SSs are:

  {\bf (iv)} angular momentum dredge-up when the convective envelope 
approaches the core region of the giant  (Simon \& Drake 1989).

  {\bf (v)} planet engulfment during the giant phase. 

The discovery of giant planets around solar type stars poses the question 
as to what happens when a MS star evolves into a giant and engulfs 
the planet (see Massaroti et al 2008 and references therein). 
Assuming for a typical M3III star R$_g=71.5$~R$_\odot$ and M$=1.6$~M$_\odot$,
we calculate that an ingestion of a planet with a mass 
0.01~M$_\odot$ can speed up the rotation of the giant by up to 40~\kms\  
(see Paper I for more details).

Some of the symbiotics belong to the old disk population (Wallerstein 1981).
Smith et al. (1996, 1997) have shown that the symbiotics 
AG~Dra and BD-21$^0$3873 contain metal poor giants.
We have selected from Carney et al. (2003, 2008a,b) 40 metal poor K giants with 
$3900\le$T$_{eff} \le 4500$ and $0.5 \le \log g \le 1.5$ and 
[Fe/H]$\le-0.8$. The statistical tests do not indicate statistically significant differences 
between symbiotic K giants and metal poor giants. 
Reasons for the fast rotation  (and/or line broadening) 
observed in the spectra of  metal poor giants
are considered inherited rotation, induced rotation due to a stellar or planetary companion, 
and more recently macroturbulence (see also Carney et al. 2003, 2008b).

Because the symbiotic stars rotate faster than the 
field giants, they probably  can help us to understand in which cases
the binarity and in which cases the planet ingestion or some other mechanism  
(i.e. inherited rotation, macroturbulence) is responsible for the fast rotation (or line broadening). 

\subsection{After-effects of the fast rotation }
Fast rotation should  enhance the mass loss in the equatorial regions,
like the disks of the classical Be stars. 
The IR-disks of D'-type symbiotics are  (probably)
a direct result of the fast rotation.                                                                    
A circumstellar medium that is significantly
denser in the equatorial regions of the binary than at the poles
is already detected in RS~Oph (Bode et al. 2007), indicating that
circumbinary disks (non-dusty) can also be formed in S-type symbiotics.

Mass-loss rates for the symbiotic giants derived from cm
and mm/submm radio observations (Seaquist, Krogulec \& Taylor 1993;
Miko{\l}ajewska, Ivison \& Omont 2003) and from IRAS data (Kenyon, Fernandez-Castro
\& Stencel 1988) are systematically higher than those reported for 
field giants. The relation between  rotation and mass  loss is not yet exactly 
quantified (see Wilkinson et al 2005).  Nieuwenhuijzen \& 
de Jager (1988) find   $\dot M \propto (v\: \sin i)^{2.46}$.
Following this relation, we estimate that if the symbiotic giants rotate 
1.5-4.0 times faster than the field giants, they should have 
3-30 times larger mass loss rates. This matches the 
observed values (see Fig.2 in Mikolajewska, Ivison \& Omont 2003) 
and means that the rapid rotation of the mass-donors 
is the likely physical reason that increases mass loss in SSs
over that exhibited by  normal giants.

Smith et al. (1996) point out that the Barium and CH stars 
are similar to symbiotics but do not trigger symbiotic like activity.
Because the mass loss and/or rotation of the mass donors 
can be the reasons for the lack of activity, 
it will be interesting to measure their rotational velocities and to compare them with 
symbiotics and field giants.

\subsection{Jet ejecting symbiotics}
Bipolar ejections (jets) are rare phenomena for the
accreting white dwarfs in symbiotics (e.g. Leedj{\"a}rv 2002). 
Among our sample of SSs  with known \vsi\ 
there are 4 objects with detected jet/blob ejection:
RS~Oph (Iijima et al. 1994, Zamanov et al. 2005),
MWC~560 (Tomov et al. 1990, Schmid et al. 2001), 
StH$\alpha$190 (Munari et al. 2001) and Hen~3-1341 (Tomov, Munari, \& Marrese 2000).

In the recurrent nova RS Oph,
the red giant probably rotates faster than the orbital period (Paper~II). 

V694~Mon (MWC~560) is seen almost pole-on. 
The  jet (orbit) inclination is $\; i<16^0$ (Schmid et al. 2001)
and  probably the orbital period P$_{orb}\approx$1931 day 
(see    Gromadzki et al. 2007 and references therein).
For a typical M6III star we assume a radius of 147.9~R$_\odot$ 
(van Belle et al. 1999).
If we assume that the rotational axis of the red giant
is perpendicular to the orbital plane, we calculate
a rotation period of the red giant P$_{rot}\approx 180$ d,
which means that $P_{rot}<<P_{orb}$. 

The rotation period of StH$\alpha$190 is P$_{rot} \le5$~d
(see Paper~I). If we assume the typical 
orbital period of a symbiotic binary P$_{orb}\ge200$~d (see Mikolajewska 2003), 
then we get $P_{rot}<<P_{orb}$. 

The rotation  period of Hen~3-1341 
is possibly P$_{rot}\approx 168$~d (assuming R$_g=57.8$~R$_\odot$, $i\approx30^0$),
which is less than the typical orbital periods of S-type SSs.

These 4 objects point to the possibility that the jet ejections 
in symbiotics are detected in systems 
where the mass donors rotate faster than the orbital period (i.e. $P_{rot}<P_{orb}$). 
In this connection  candidates for detection of such ejections 
should be considered among the fast rotating SSs V840~Cen, Hen~3-1213, Hen~3-1674, as well 
as all D'-type symbiotics.


\section{Conclusions}

In this paper we report measurements of the projected rotational velocity
(\vsi) of the cool giants in  9 symbiotic stars and 28 field giants,
using high resolution spectroscopic observations. 
Collecting  together all available data (from Paper I,II,III  as well as from the literature),
we compare the rotation of the mass donors in symbiotic stars
with the rotation  of field  giants of similar spectral types
and find  that:

(1) The symbiotic  K giants included in our survey 
rotate on average more than twice as
fast as the field K giants. 

(2) The  M giants in symbiotics  
are more rapid rotators  than most of the field M giants.

(3) For  the D'-type (yellow) symbiotics, 5 out of the six southern D'-type SSs 
are the fastest or among the fastest rotators in their spectral classes. 

(4) As a result of the rapid rotation 
SSs should have on average $\sim\!$10 times larger mass loss rates
than normal red giants.

(5)  We find suggestions that in the jet-ejecting symbiotics 
the mass donors  rotate faster than the orbital periods.

This is the first observational investigation
that clearly confirms theoretical predictions 
that the mass donors in symbiotics are fast rotators.

\section*{Acknowledgments}  
We acknowledge stimulating discussions with the late Dr. John M. Porter.  
This research has made use of SIMBAD, IRAF, and  MIDAS.  
IKS acknowledges the partial support from Bulgarian NSF grant F-1403/2004. 
RKZ was PPARC research assistant and MFB was supported by a PPARC/STFC Senior Fellowship 
during the early stages of this work.
The authors are grateful to an 
anonymous referee for valuable comments on earlier versions of this paper.


\end{document}